\pdfoutput=1
\documentclass[preprint,floats,floatfix,aps,epsfig,nofootinbib,amssymb]{revtex4}
\usepackage{subfigure}
\usepackage{hyperref}
\usepackage{graphics,psfrag}
\usepackage{graphicx,psfrag}
\usepackage{amsmath,amssymb}
\usepackage[sort&compress]{natbib}
\usepackage{cancel}
\usepackage{bm}
\usepackage{verbatim}
\usepackage{multirow}
\usepackage{ulem}
\usepackage{color}

\def\beq{\begin{equation}}
\def\eeq{\end{equation}}
\def\bea{\begin{eqnarray}}
\def\eea{\end{eqnarray}}

\begin{document}
%

\preprint{
   {\vbox {
      \hbox{\bf MSUHEP-130201}     
      }}}
\vspace*{0.5cm}

\title{{Searches for New Vector Like Quarks: Higgs Channels}}
\vspace*{0.25in}   
\author{{Anupama Atre$^{1}$\footnote{{\tt Email: avatre@pa.msu.edu}.}, Mikael Chala$^{2}$\footnote{{\tt Email: miki@ugr.es}.} and Jos\'e Santiago$^{2}$\footnote{{\tt Email: jsantiago@ugr.es}.} }}
\affiliation{\vspace*{0.1in}
$^1$ Department of Physics and Astronomy, Michigan State University, East Lansing, MI 48824, U.S.A.\\
$^2$ CAFPE and Departamento de F\'{\i}sica Te\'orica y del Cosmos, University of Granada, 18071 Granada, Spain\\}
\vspace*{0.25 in} 


\begin{abstract}
\vspace{0.5cm}
\noindent
New vector-like quarks can mix sizeably with first generation Standard
Model quarks without conflicting with current experimental
constraints. Searches for such new quarks have been performed in pair
production and electroweak single production channels with subsequent
decays into electroweak gauge bosons. To fully explore the underlying
structure of the theory the channels with heavy quark decays into
Higgs bosons are crucial and in this article we consider for the first
time the LHC reach for such channels.  
The two main production mechanisms involve single
production of new quarks through the fusion of a vector boson and
the Higgs and single production in association with a Higgs
boson. We show that both channels have promising reach at
the LHC and that they complement the current direct searches involving 
decays into electroweak gauge bosons.

\end{abstract}

\maketitle


\section{Introduction}

New vector-like quarks are a common ingredient in models of new
physics. The mixing of new vector-like quarks with Standard Model (SM)
quarks  induces corrections to the SM quark couplings that are
proportional to the square of the 
mixing angles ~\cite{delAguila:1982fs,delAguila:2000rc}. 
This has lead to the misconception that
vector-like quarks cannot mix sizeably with first generation SM
quarks, given the stringent constraints on SM quark couplings from
precision flavor and electroweak (EW) observables. Motivated by models
with warped extra dimensions~\cite{Carena:2006bn,Carena:2007ua} 
it was pointed out
in Ref.~\cite{Atre:2008iu} that when first generation SM quarks mix
with several new quarks with 
different 
quantum numbers, the contributions of the different multiplets to the
anomalous SM quark couplings can cancel among themselves, thus leaving no low
energy trace even for large values of mixing. These cancellations can in fact
be enforced by a symmetry~\cite{Agashe:2006at} naturally present in composite
Higgs models with minimal flavor
violation~\cite{Redi:2011zi}. Indirect constraints provide only mild
restrictions in these models and the direct searches advocated 
in Refs.~\cite{Atre:2008iu,Atre:2011ae} and recently performed
in Refs.~\cite{Aad:2011yn,atlassingle} 
become the main probes of these new vector-like quarks. 

The prototypical example in which these cancellations are present 
is the degenerate bidoublet model. It contains
four new quarks, two with electric charge $2/3$, one with electric
charge $5/3$ and the last one with electric charge $-1/3$. 
These new quarks have a unique decay mode to the up quark 
and the $Z$, $H$, $W^+$ and $W^-$ bosons, respectively,
all with 100$\%$ branching fraction. Current direct searches involve
only decay modes involving EW gauge bosons and are therefore sensitive
to three out of the four quarks in the model. 
In this article we study for the first time the ability of the Large Hadron Collider (LHC) to
measure the Higgs channels in the degenerate bidoublet 
model. These channels are crucial to fully test the nature of the
model. Higgs channels relying on the presence of new particles beyond
the heavy quarks (a massive color octet vector boson) have been
recently studied in the degenerate bidoublet model in the context of
composite Higgs models in Ref.~\cite{Carmona:2012jk}. In this article we
consider the minimal case in which the four new quarks are the only
new particles at the relevant energies.

New quarks can be pair or singly
produced. Pair production is dominated by QCD interactions and is
therefore model-independent. It is however not directly sensitive to
the coupling between the new quarks and the SM quarks but only to the
corresponding branching fractions. Furthermore, for sizeable values of
the mixing angles that we are interested in single production is
actually a more sensitive probe of these models as shown in Refs.~\cite{Atre:2008iu,Atre:2011ae} and also by current
experimental limits in each channel~\cite{Aad:2012bt,atlassingle}.  
Thus, we 
disregard pair production in this article and focus on single
production. Single production can proceed via different mechanisms. The most relevant ones in the model under consideration
are $t$-channel exchange of the Higgs boson, single production through
fusion of the Higgs and an EW gauge boson and single production in
association with a Higgs boson. We will show that the latter two
production mechanisms have a good reach at the LHC and are capable of providing an important test of the model if it is realized in
nature. Given current constraints from single production in the EW
gauge boson channels we focus on the LHC running at its highest
designed luminosity with $\sqrt{s}=14$ TeV.  

The rest of the article is organized as follows. We describe the model
and its main features in Section~\ref{sect:model}. We then discuss the
main processes involving the Higgs channels in the model in
Section~\ref{sect:production} and dedicate
Sections~\ref{sect:vbfusion} and~\ref{sect:associated} to the detailed
analysis of the LHC reach in the two most promising channels:
vector-boson fusion and associated production. Finally we discuss our
results and conclude in Section~\ref{sect:conclusions}.

\section{Degenerate Bidoublet Model: Definition and
  Constraints \label{sect:model}} 

\begin{figure}[t]
{
\includegraphics[width=0.49\textwidth, clip=true]{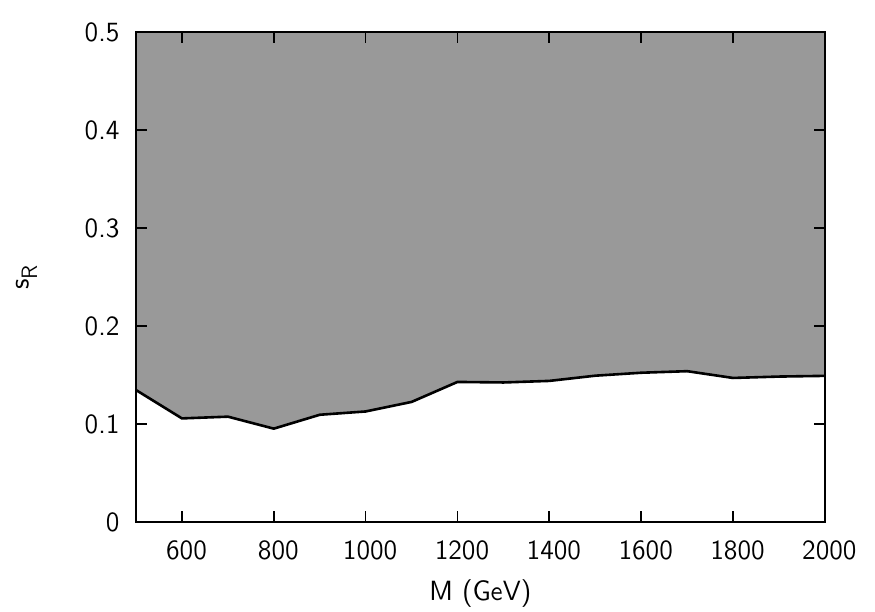}
\hfil
\includegraphics[width=0.49\textwidth, clip=true]{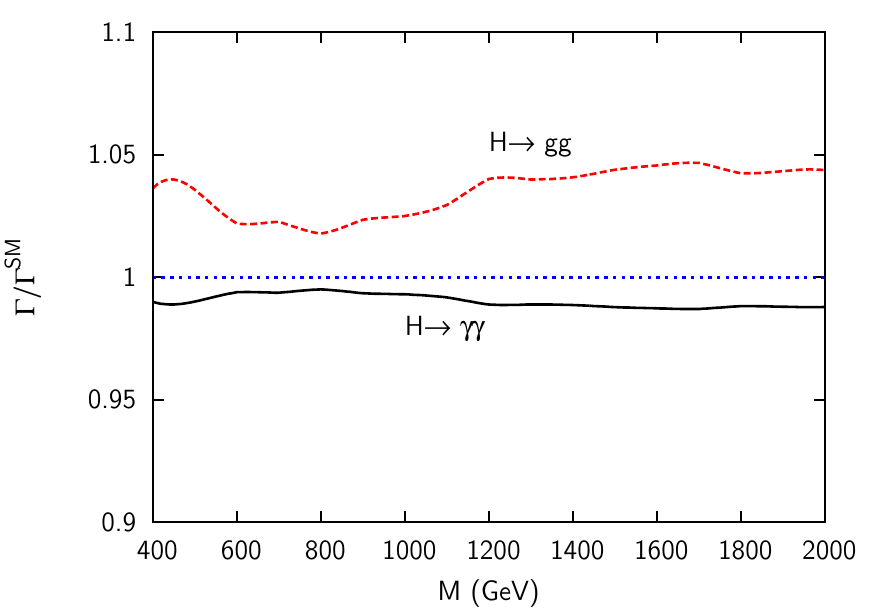}
}
\caption{(a) Left: upper
bound on the value of the mixing angle, $s_R,$ as a function
of the common heavy quark mass, $M,$ for 
  the degenerate bidoublet derived from direct
  searches~\cite{atlassingle}; (b) right: 
Maximum deviation of Higgs decay widths $\Gamma(H\to
gg,\gamma\gamma)$ with respect to the SM in the degenerate bidoublet
model when the bounds from direct searches (left panel) are taken into account. 
}
\label{fig:sRdirectbounds}
\end{figure}

The degenerate bidoublet model has been described in full detail
in Refs.~\cite{Atre:2008iu,Atre:2011ae} (for an implementation in the
lepton sector see Ref.~\cite{delAguila:2010es,Carmona:2013cq}). 
It consists of two new vector-like
doublet quarks of hypercharges $1/6$ and $7/6$, respectively, with
identical Dirac mass, that mix with equal strength  
only with the up quark in the basis in which all the non-trivial
flavor structure resides in the down sector. The spectrum 
includes two charge $2/3$ quarks (denoted by $U_Z$, $U_H$), 
one of charge $5/3$
(denoted by $X$) and one of charge $-1/3$ (denoted by $D$). All observables can be
parameterized in terms of two masses, that we take to be
$m_u$, the SM up quark mass and the 
common mass $M=m_{U_Z}=m_D=m_X$, and a mixing angle $s_R\equiv \sin
\theta_R$. The relevant relations between these parameters are
\begin{eqnarray}
&& m_{U_H}= M c_L/c_R, \quad s_L=s_R m_u^2/M^2\approx 0, \nonumber \\
&&\kappa^R_{uU_Z}=\sqrt{2}\kappa^R_{uX}=\sqrt{2}\kappa^R_{uD}= s_R,
\quad y^R_{uU_H}=s_R c_R \frac{m_{U_H}}{v}.
\end{eqnarray}
All other couplings are either vanishing or suppressed by powers of
$s_L$ and therefore negligible.
From these couplings it should be clear from the choice of notation, the
only allowed decays for the charge $2/3$ quarks are  $U_Z\to Z u$ and
$U_H\to H u$. Similarly, $X$ and $D$ can only have charged current decays.
This model enjoys a custodial symmetry protection of the SM quark
gauge couplings. They only receive corrections proportional to powers
of $|s_L|\lesssim 10^{-11}$ from the mixing with the new quarks.
One loop 
constraints give the very mild bound~\cite{Atre:2011ae}
\begin{equation}
|s_R|\lesssim 0.75~.
\end{equation}

Direct searches result in the most stringent constraints for the
degenerate bidoublet model. We have translated the constraints
obtained in Ref.~\cite{atlassingle} to the degenerate
bidoublet case 
and the results are shown in Fig.~\ref{fig:sRdirectbounds}(a). In Ref.~\cite{atlassingle} the bounds were obtained assuming only one type of quark at a time. In our model we have only one quark, $U_Z$, contributing to the neutral current channel, and therefore the experimental bound as given in Ref.~\cite{atlassingle} applies to our analysis as well. However for the charged current channel we have two quarks, $X$ and $D$, contributing simultaneously whereas in Ref.~\cite{atlassingle} only one quark was considered. Hence in extracting the limits on the coupling we have to consider the case that both quarks are simultaneously present. The explicit expressions for including different types of quarks are given in Ref.~\cite{Atre:2011ae}. The bound presented in Fig.~\ref{fig:sRdirectbounds}(a) is the most stringent one of the charged and neutral current channels.
These constraints are stringent enough that Higgs searches do not
impose any further constraints. For instance, given the current bounds
in Fig.~\ref{fig:sRdirectbounds}(a), gluon fusion is enhanced with
respect to the SM by less than $5\%$ and the $h\to \gamma \gamma$
channel decreased by less than $2\%$ as shown in
Fig.~\ref{fig:sRdirectbounds}(b). These direct constraints on $s_R$
also 
imply that $m_{U_H}\approx M$ with an approximate precision of $1\%$,
well within the experimental resolution. Thus, from now on we will
consider all four quarks to be degenerate.

\section{Higgs Production Channels\label{sect:production}}

\begin{figure}[tb]
{
\includegraphics[width=0.27\textwidth, clip=true]{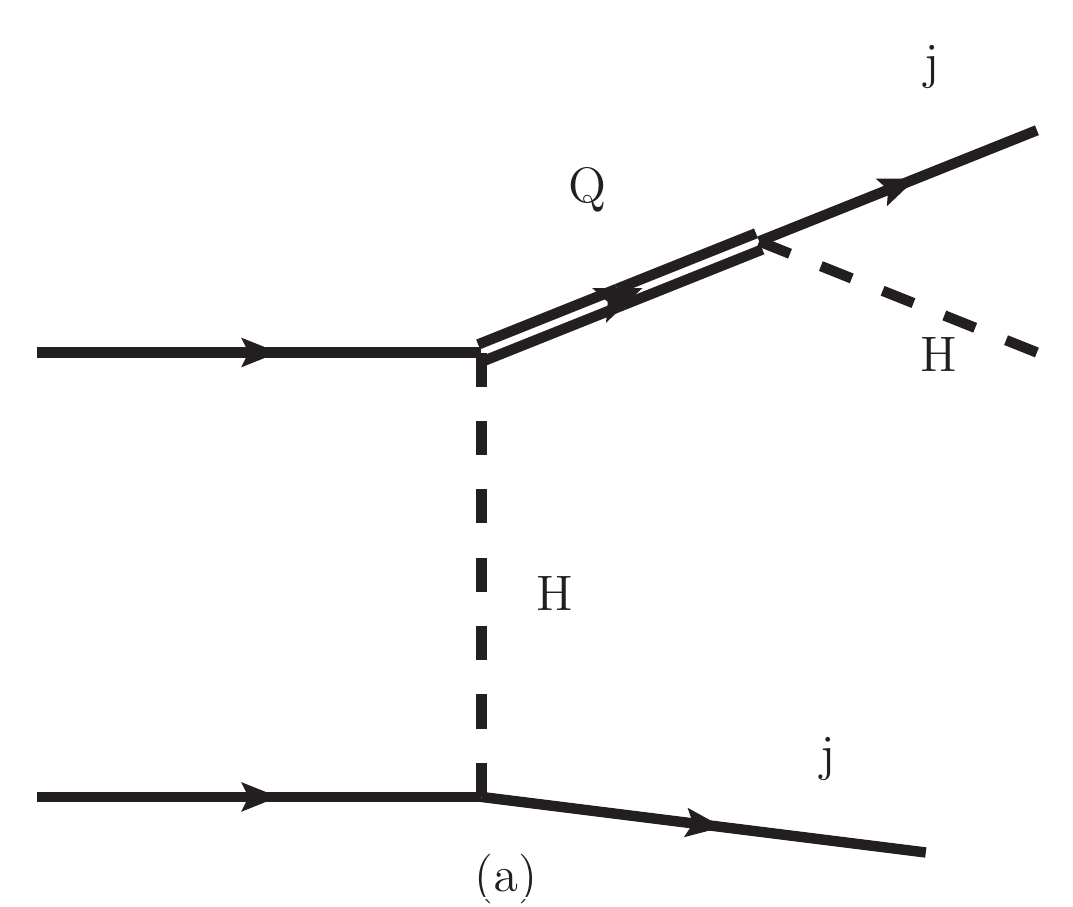}
\hfil
\includegraphics[width=0.27\textwidth, clip=true]{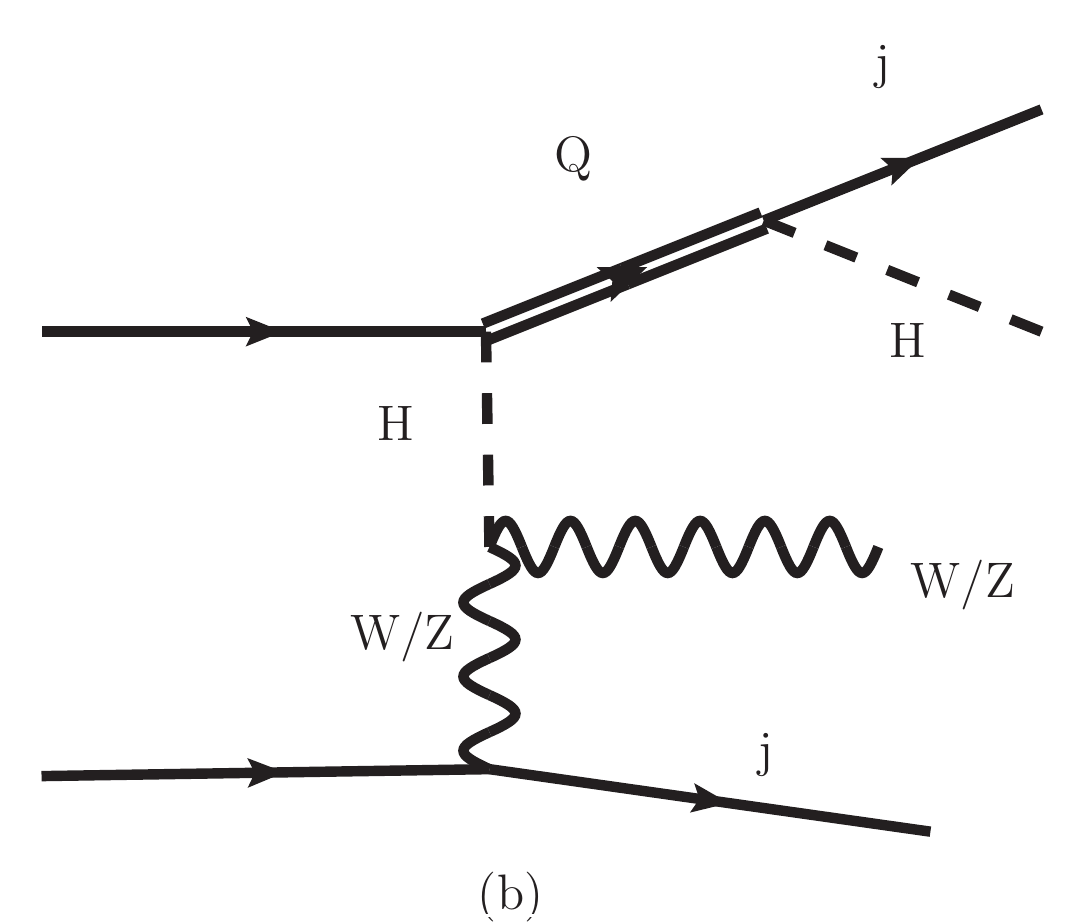}
\hfil
\includegraphics[width=0.27\textwidth, clip=true]{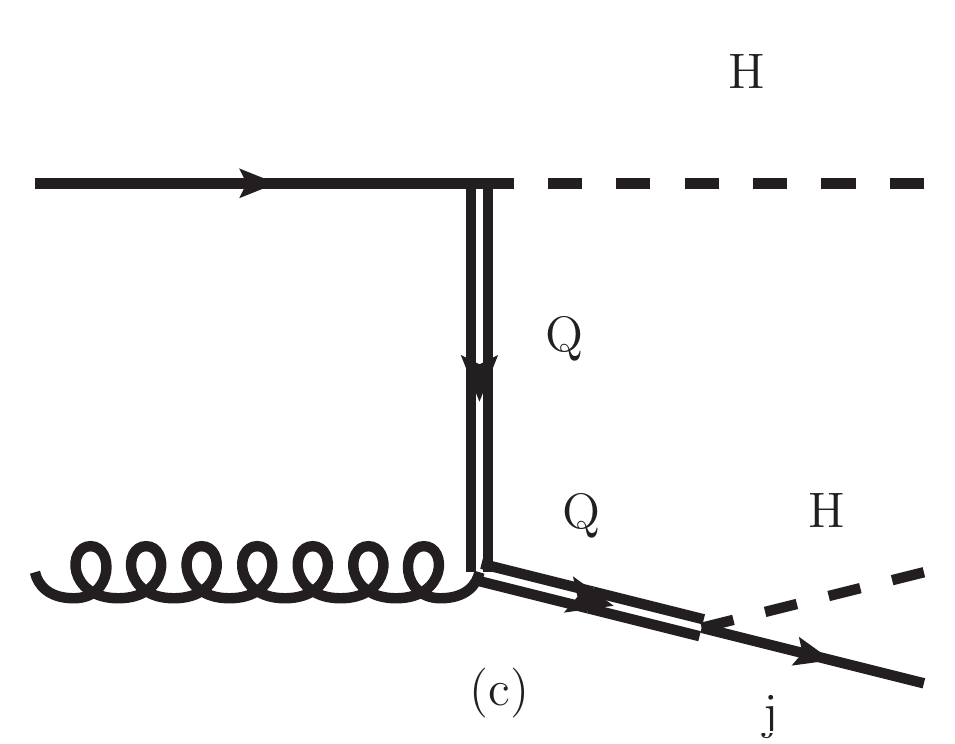}
}
\caption{Sample diagrams for: (a) single production, 
(b) vector boson Higgs fusion and (c) associated production  
of $U_H$ with subsequent decay into $H u$. 
}
\label{fig:production:diagrams}
\end{figure}

\begin{figure}[tb]
{
\includegraphics[width=0.7\textwidth, clip=true]{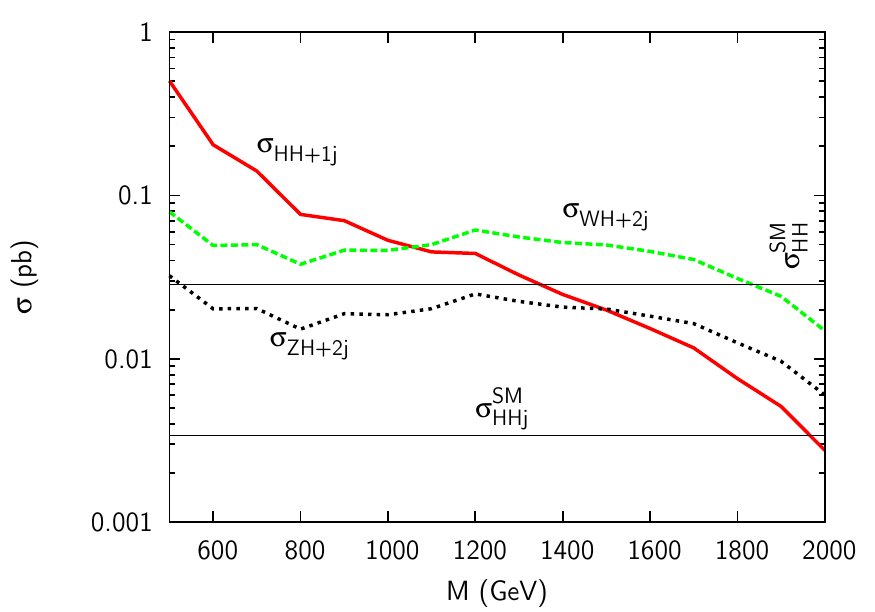}
}
\caption{
Cross sections (in pb) for the associated production channel (labeled $HH$+$1j$)
and the vector boson Higgs fusion channel 
(labeled $WH$+$2j$ or $ZH$+$2j$) for $s_R$ fixed to the current
upper bound. 
For reference we also show the SM production cross section of two
Higgs bosons and two Higgs bosons plus a hard jet with
$p_T(j)>100$ GeV. 
}
\label{fig:production}
\end{figure}

We describe in this section the relevant single production
mechanisms of new quarks that
involve at least one Higgs boson in the final state in the context of
the degenerate bidoublet model. 
One important feature of the model is the fact that each of the heavy quarks couples to the $u$ quark and only
one gauge boson or the Higgs, {\it i.e.} $Br(X \to uW^+) = Br(U_Z \to
uZ) = Br(D \to uW^-) = Br(U_H \to uH) = 100\%$. This strongly restricts the number of
relevant diagrams that contribute to Higgs production through the
decay of the heavy quark $U_H$. The three mechanisms that we consider
and their relevant features are: 
\begin{itemize}
\item {\it Single production: } $qq' \to jU_H \to jjH$.\\
 In this channel the heavy quark, $U_H$, is produced in association with a single (forward) jet. Subsequent decay of the heavy quark to a jet and the Higgs boson leads to a final state of $Hjj$ as shown in Fig.~\ref{fig:production:diagrams}(a). Single production is suppressed in our model by the up quark Yukawa coupling and will be disregarded henceforth.
 
\item {\it Vector boson Higgs fusion} (VBHF:) $qq' \to jVU_H \to
jjVH$, where $V=W,Z$.\\  
In this channel the heavy quark, $U_H$, is produced in association
with a (forward) jet and an EW gauge boson. After the decay of the
heavy quark we have a final state of $VH$+$2j$, where $V=W,Z$ gauge
boson as shown in 
Fig.~\ref{fig:production:diagrams}(b). The VBHF production mechanism is
initiated by two valence quarks, involves unsuppressed couplings and
has a 
longitudinal gauge boson enhancement. Thus, the corresponding cross
section can be sizeable and relatively insensitive to the mass of the
heavy quark. This is shown in  
Fig.~\ref{fig:production} where the cross sections correspond to
the currently allowed values of $s_R$.  
The presence of an EW gauge boson allows for a clean trigger
using its leptonic decays thereby allowing the use of the dominant
$b\bar{b}$ Higgs decay.  

\item {\it Associated Production:} $qg \to HU_H \to jHH$.\\ 
In this channel the heavy quark, $U_H$ is produced in association with
a Higgs boson and the subsequent decay of the heavy quark leads to a
unique two Higgs plus a hard jet final state shown in 
Fig.~\ref{fig:production:diagrams}(c). Double Higgs production has been
studied as a way of measuring Higgs
self-couplings~\cite{Baur:2003gp,Dolan:2012rv,Goertz:2013kp} and anomalous Higgs
couplings~\cite{Grober:2010yv,Contino:2012xk,Dolan:2012ac}. The
presence of a hard jet from the decay of the heavy quark in our
analysis enhances signal over background. Associated production is
initiated by a valence quark and a 
gluon. Hence the cross sections can be quite large for small
values of $M$ but suffer a stronger suppression for larger values
of the heavy quark mass, due to the steeply falling gluon parton distribution
functions (PDFs). We show the
cross section for this process in Fig.~\ref{fig:production} for the
currently allowed values of $s_R$.   
For comparison we also show in Fig.~\ref{fig:production}
 the production cross section for $HH+X$ and $HH+j+X$ in the SM, with
 $p_T(j)\geq 100$ GeV, as computed in~\cite{Dolan:2012rv}. 
\end{itemize}

In the following two sections we describe the analyses we
propose to measure the VBHF and associated production mechanisms. We have generated
signal and background events at partonic level using \texttt{Madgraph}
v 4.5.0~\cite{Alwall:2007st} and \texttt{ALPGEN}
v2.13~\cite{Mangano:2002ea}, respectively.  
We have then used
\texttt{Pythia} v 6.4~\cite{Sjostrand:2006za} 
for hadronization and showering and
\texttt{Delphes} v 1.9~\cite{Ovyn:2009tx} for 
fast detector simulation. The detector card in \texttt{Delphes} has been
modified to better agree with published
distributions~\cite{Hubaut:2005er,Hubaut:2006zxa,Castro:1135130}. The main 
changes are: the tracking efficiency was updated to 95; the isolation
criterion was changed to $\Delta R=0.4$ and a value of 0.7 was used
for the b-tagging efficiency. We have used the CTEQ6L1
PDFs~\cite{Pumplin:2002vw} and use $M$ for
factorization and renormalization scales for signal and the default values for
background. Thus, $\sum m_Q^2+p_T^2$ was used for $Q\bar{Q}$+jets and $Q\bar{Q}Q\bar{Q}$+jets (with $Q = b,t$), where the sum extends to all the final state partons, while $m_V^2+p_{T,V}^2$ was used for $V$+jets (with $V = W,Z$). Jets and leptons are defined as requiring $p_T^j > 30$
GeV, $|\eta_j | <  5$, $p_T^\ell > 20$ GeV, $|\eta_\ell| < 2.5$ and
$\Delta R(j\ell) > 0.4$, 
where jets are clustered using the anti-$k_T$ algorithm using a cone radius of
$\Delta R=0.4$. Throughout the analysis we have used the Higgs mass
to be $m_H=125$ GeV. We show in Table~\ref{tab:bkgxsec} the  
relevant backgrounds for the analyses in this article with their
corresponding cross sections. 

\begin{table}[t]
\begin{tabular}{|l|c|l|c|}\hline
{\centering Background} & $\sigma (pb)$ & {\centering Background} & $\sigma (pb)$\\
\hline
\ $t \bar{t}$ semileptonic $+\ 0 - 4 j$ &  222.3 &
\ $W_{\ell\nu}WW + 0 - 2 j$ & 0.14 \\
\ $t \bar{t}$ fully leptonic $+\ 0 - 4 j$  &  54.3 &
\ $WWZ + 0 - 2 j$ & 0.18 \\
\ $W_{\ell\nu} b\bar{b} + 0 - 2 j$ &  12.5 &
\ $W_{\ell\nu} + 1j,\ p_T^{j_h} > X$ &  178.1 \\
\ $b\bar{b}b\bar{b}+ 1j,~p_T^{j_h}>X$ & 3.1 &
\ $t \bar{t} b \bar{b}$ & 0.85 \\
\ $b\bar{b}+3j$ &  515.0 &
\ $W_{\ell\nu} W + 0 - 2 j$ & 49.0 \\
\ $Z_{\ell\ell}/\gamma_{\ell\ell} + 1 - 4j$ &  552.7 &
\ $WZ + 0 - 2 j$ &  39.9 \\
\ $Z_{\ell\ell} Z + 0 - 2 j$ & 2.37  &
\ $Z_{\ell\ell}b\bar{b}+0-2j$ &  4.5\\
\hline
\end{tabular}\hspace{0.6cm}
\caption{Cross sections (in pb) for the various background processes. In our
  notation $W_{\ell\nu}$ and $Z_{\ell\ell}$ represent leptonic decays
  of the $W$ and $Z$ gauge bosons and $\ell=e,\mu,\tau$. The transverse
  momentum cut for the hardest jet ($j_h$) for W+jets background is
  $X=130$ GeV. 
In the case of $b\bar{b}b\bar{b}+1j$ we have $X=150$
  GeV. 
The explicit number of jets listed
  stands for the ones generated at the parton level and the rest of
  the jets are from initial and final state radiation. The first three are the main backgrounds to the VBHF channel and the next two are the main ones for the 4b+j associated production channel. The other background processes listed in the table have been considered in the analysis but become irrelevant after all the optimization cuts have been applied.
} 
\label{tab:bkgxsec}
\end{table}

\section{Vector Boson Higgs Fusion \label{sect:vbfusion}}

In the vector boson Higgs fusion channel the heavy quark,
  $U_H$, is produced singly in association with a $W$ or $Z$ gauge
  boson and a jet leading to the final state
 \begin{equation}
pp\to V U_H j \to  V H j j,
\end{equation}
where $V=W,Z$. Of the two jets in the final state, the one coming from
the heavy quark decay tends to be quite hard whereas the other one
tends to be relatively forward. Furthermore the Higgs
boson comes from the decay of a massive particle ($U_H$) and is
typically quite boosted. These features can be used to enhance signal
over backgrounds. Considering the leptonic decays of the gauge boson
helps reduce QCD backgrounds as well as provide a clean trigger. Hence
we will consider only the leading $H\to b\bar{b}$ decay.  
The final state is therefore $b\bar{b} jj \ell \cancel{E}_T$ or $b\bar{b}
jj \ell\ell$, for $V=W$ or $Z$ gauge boson respectively. 
The latter process
is potentially cleaner but suffers from reduced statistics. The cross
section is of ${\cal O}(fb)$ once the decay branching fractions of
the $Z$ and $H$ are included. 
Thus, even a minimal set of cuts quickly reduces the
number of events to just a few except for very large luminosities. For
this reason we focus on the more promising charged
current channel.

\subsection{Charged current channel}

For the charged current channel with the $b\bar{b} jj
\ell \cancel{E}_T$ final state we have implemented the
following cuts:
\begin{itemize}
\item Particles: exactly one charged lepton ($e$ or $\mu$) plus at
least four jets with 
  exactly two b tags.
\item Hard jet: $p_T(j_h)\geq 200$ GeV for the hardest jet that is not 
b-tagged.
\item Higgs reconstruction: $|m_{bb}-m_H|\leq 30$ GeV.
\item Higgs boost: $\Delta R(bb)\leq 1$.
\item Heavy quark mass: $|m_{bbj_h} -M|\leq 200$ GeV. 
\end{itemize}

\begin{figure}[htb]
{
\includegraphics[width=0.45\textwidth, clip=true]{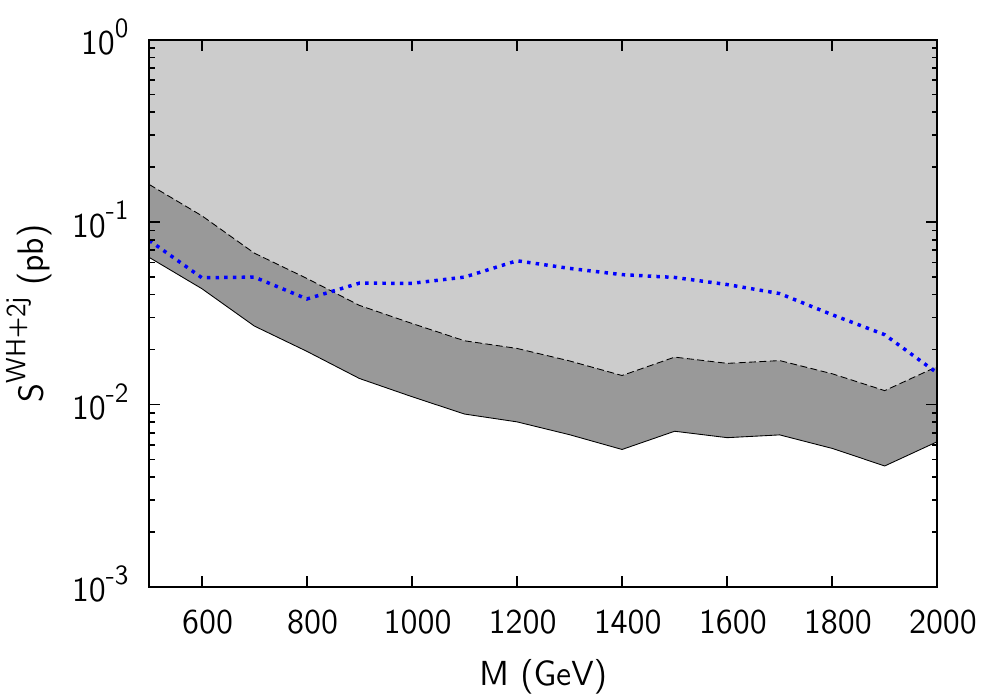}
\hfil
\includegraphics[width=0.45\textwidth, clip=true]{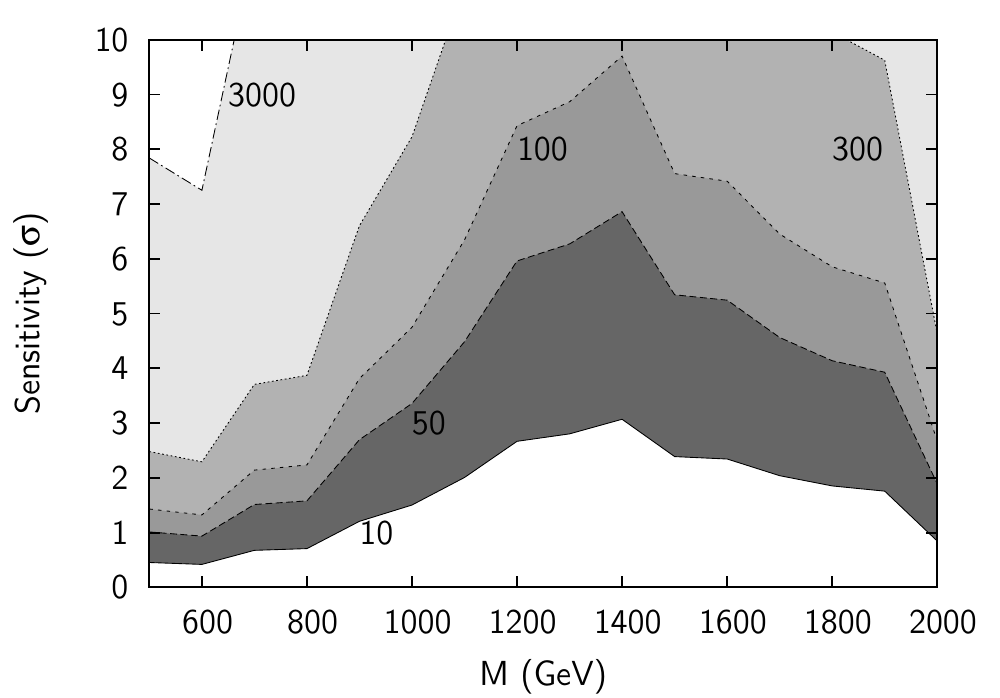}
}
\caption{(a) Left: 2 and 5$\sigma$ bounds (indicated by dark and light colored regions respectively) on the production cross section
  times branching fraction for the $WH jj$ channel as a function of the heavy quark mass, $M$, for
  the LHC with $\sqrt{s}=14$ TeV and an integrated luminosity of 300
  fb$^{-1}$. For reference we also show the current (95$\%$ C.L.) 
upper bound (indicated by the dotted blue curve) in the
  context of the degenerate bidoublet model. (b) right: contour plot
  of the luminosity required for a certain degree of confidence level
  as a function of the heavy quark mass in the degenerate bidoublet
  model with $s_R$ fixed to the current limit.}
\label{fig:sensitivityWF}
\end{figure}

We show in Table~\ref{tab:WFeffs} the signal cross section and the
total efficiencies for the 
signal and main backgrounds for different values of the heavy quark
mass. All relevant background processes have been considered but we
only report the ones that are non-negligible after all
the cuts in Table~\ref{tab:WFeffs}.
In Fig.~\ref{fig:sensitivityWF}(a) we show the 2 and 5$\sigma$
sensitivity for the production
cross section times branching fraction for the $WHjj$ channel as a
function of the heavy quark mass, 
$M$, for the LHC with $\sqrt{s}=14$ TeV and an integrated
luminosity of 300 fb$^{-1}$. For reference we also show the current (95$\%$ C.L.) 
upper bound (indicated by the dotted blue curve) in the
  context of the degenerate bidoublet model. In Fig.~\ref{fig:sensitivityWF}(b) 
we show the confidence
level that can be reached, as a function of $M$ for different
values of total integrated luminosity when $s_R$ is fixed to the
current upper limit.  As seen in Fig.~\ref{fig:sensitivityWF} new regions in parameter
space that are currently unconstrained can be probed in the early runs
with 10 (50)fb$^{-1}$ at 
2 (5)$\sigma$ sensitivity. The shape of the contours in these figures
can be understood from the production cross section, which is flat as a
function of $M$ as shown in Fig.~\ref{fig:production}. 
Larger values of $M$ give rise to a similar number of signal
events but the requirement on the heavy quark mass reconstruction reduces the background more efficiently for higher masses of the heavy quark. For
$M\gtrsim 1.5$ TeV the production cross section decreases due to the decreased parton luminosity and
so does the sensitivity.

\begin{table}[t]
\begin{tabular}{|l|c|c|c|c|}\hline
M (GeV) & $\sigma_s$ (fb) & $\epsilon_s$ & $ \epsilon_{t\bar{t}}$ &
$\epsilon_{Wb\bar{b}}$ \\\hline
500 & 79 & 0.010 & $1.0\times 10^{-4}$ & $1.4\times 10^{-4}$ \\
1000 & 46 & 0.040 & $4.7\times 10^{-5}$ & $7.7\times 10^{-5}$ \\
1500 & 50 & 0.025 & $7.7\times 10^{-6}$ & $1.4\times 10^{-5}$ \\
\hline
\end{tabular}\hspace{0.6cm}
\caption{Cross sections (in fb) for the signal ($WH jj$) 
and efficiencies for
  signal and main backgrounds for different values of $M$. The corresponding background cross sections are listed in Table~\ref{tab:bkgxsec}.
}\label{tab:WFeffs}
\end{table}

\section{Associated Production\label{sect:associated}}

In the associated production channel, the heavy quark, $U_H$, is
produced in association with a Higgs boson as shown in Fig.~\ref{fig:production:diagrams}(c). Subsequent decay of the heavy
quark leads to the unique final state with two Higgs bosons and a hard
jet: 
\begin{equation}
pp \to HU_H \to HHj.
\end{equation}
Double Higgs production has received some attention as a means of measuring the
Higgs self-couplings in the SM, see for
instance Refs.~\cite{Baur:2003gp,Dolan:2012rv,Goertz:2013kp}. In
the case of the SM, this is a very difficult 
measurement at the LHC due to the very low cross section,
$\sigma_{SM}(pp\to HH+X)=28.4$ fb~\cite{Dolan:2012rv}. As shown in 
Fig.~\ref{fig:production} the double Higgs production cross section can be
larger (by up to an order of magnitude for the lowest masses) 
in our model for masses $M\lesssim 1.4$ TeV. Furthermore, 
the presence of a very hard jet from the decay of the heavy
quark in our model provides better signal sensitivity over
background. In fact, even for the lowest value of the masses
considered, 
in over 90 $\%$ of the signal events the hardest jet has $p_T(j_h)\geq
100$ GeV whereas in the SM the corresponding cross section goes down
to $\sigma_{SM}(p\to HHj+X;p_T^j\geq 100\mbox{ GeV})=3.2$ fb.
Even without the extra hard jet it has been
argued recently that 
the LHC can be sensitive to double Higgs production in models beyond the
SM with anomalous Higgs
couplings~\cite{Grober:2010yv,Contino:2012xk,Dolan:2012ac} with an
enhancement factor with respect to the SM cross section similar to the
one present in our model. Hence our analysis which probes the underlying structure of the degenerate bidoublet model is competitive with the other double Higgs studies (which probe different theoretical aspects). 

The presence of two Higgs bosons can lead to different final states
based on the decay modes of the Higgs. We will consider two scenarios:
one where both Higgs bosons decay to $b \bar b$ and the other where
one Higgs boson decays to $b\bar b$ and the other to a pair of
photons. This gives rise to the final states $b\bar b b \bar b j$ and
$b\bar b \gamma \gamma j$ respectively.  
In the following sections we will estimate the LHC
reach for  the associated 
production channel with the two final states mentioned above 
in the context of the degenerate bidoublet model. 

\subsection{$b\bar{b}\gamma \gamma$ channel}

\begin{figure}[htb]
{
\includegraphics[width=0.45\textwidth, clip=true]{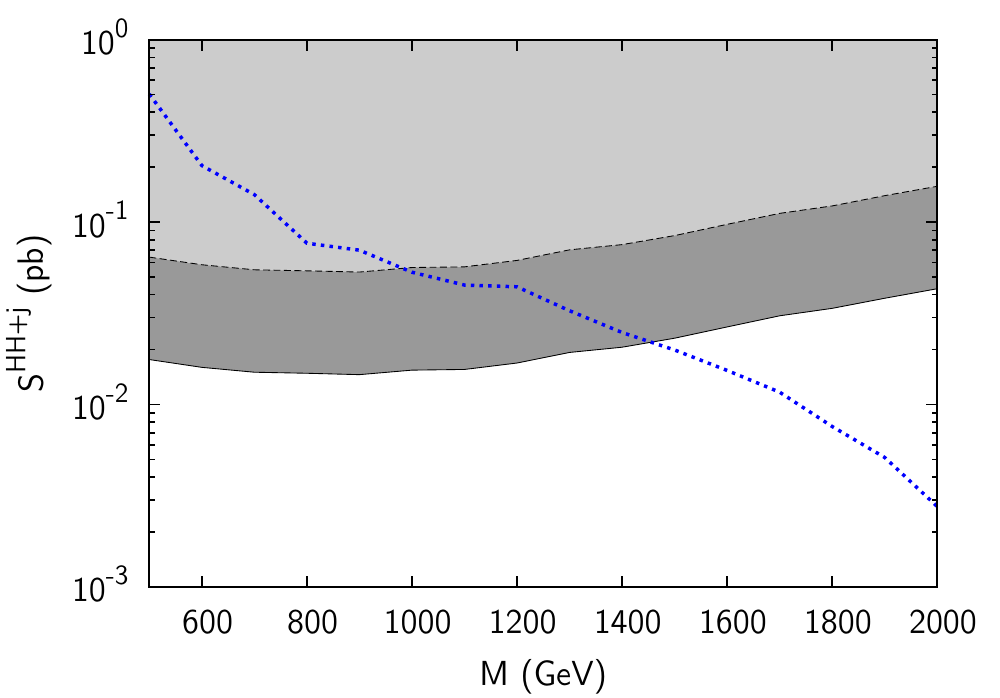}
\hfil
\includegraphics[width=0.45\textwidth, clip=true]{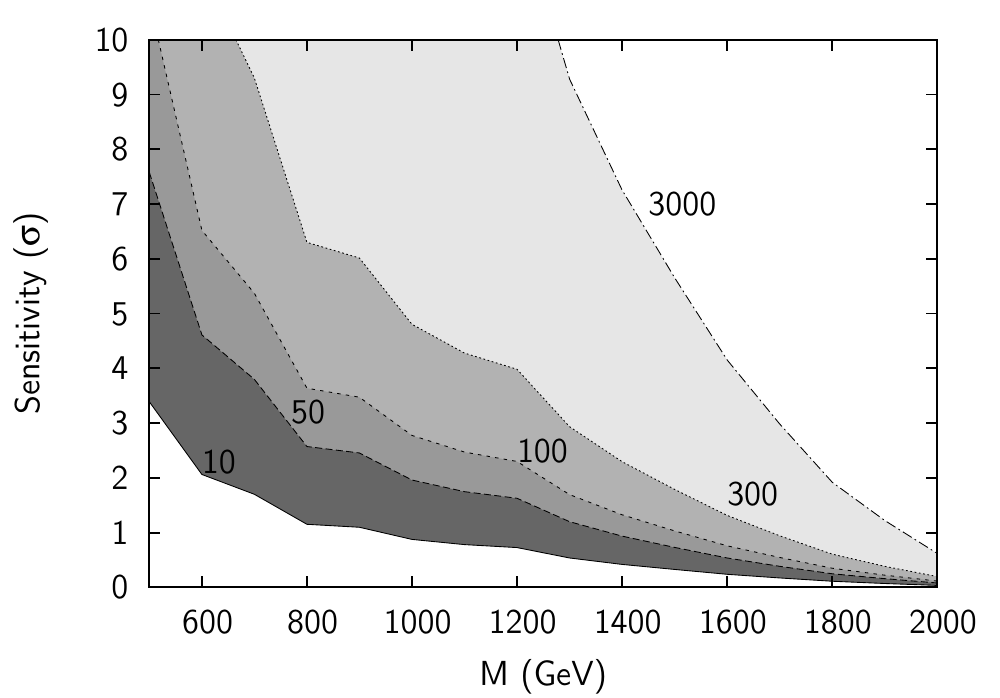}
}
\caption{
Same as Fig.~\ref{fig:sensitivityWF} but for the $HHj$ channel with
  $b\bar b \gamma\gamma j$ final state.} 
\label{fig:bbgammagamma}
\end{figure}

The $b \bar b\gamma\gamma$ channel has been studied in detail in
Ref.~\cite{Baur:2003gp} in the context of the SM and
in Ref.~\cite{Contino:2012xk} in models with anomalous di-Higgs couplings
like composite Higgs models. The result is that the cross section is
too small for a reasonable measurement in the SM but in some composite
Higgs models one could reach discovery in this channel for values of
the cross section 
$\sigma(pp\to HH \to b\bar{b}\gamma \gamma) 
\gtrsim 6 \times \sigma(pp\to HH \to b\bar{b}\gamma
\gamma)^{SM}$~\cite{Contino:2012xk}. 
In our model there are regions of parameter space in which the
double Higgs production cross section is enhanced by an even larger factor with
respect to the SM, even before taking into account the presence of a
hard jet. In order to estimate the LHC reach we have
generated events for the signal and the irreducible background. We
have implemented the following cuts as suggested in
Ref.~\cite{Baur:2003gp}:
\begin{itemize}
\item Two b-tagged jets and two photons satisfying
\begin{eqnarray}
&&p_T(b)>45\mbox{ GeV}, \quad |\eta(b)|<2.5,\quad \Delta R(b,b)>0.4,
  \nonumber \\
&& p_T(\gamma)>20\mbox{ GeV}, \quad |\eta(\gamma)|<2.5,\quad \Delta R(\gamma,\gamma)>0.4,
\label{cut1:bbgammagamma}
\\
&&|m_{bb}-m_H|<20\mbox{ GeV},\quad |m_{\gamma\gamma}-m_H|<2.3\mbox{ GeV},\quad 
\Delta R(\gamma,b)>0.4~.\nonumber 
\end{eqnarray}
\item Angular cuts
\begin{equation}
\Delta R(b,\gamma)>1.0,\quad
\Delta R(\gamma,\gamma)<2.0~.
\label{cut2:bbgammagamma}
\end{equation}
\end{itemize}
Finally we also impose an extra cut on the $p_T$ of the hardest jet to
further reduce the background:
\begin{equation}
p_T(j_h)> 100\mbox{ GeV}.
\label{cut3:bbgammagamma}
\end{equation}
The signal and irreducible
background efficiencies for the 
cuts in Eq.~(\ref{cut1:bbgammagamma}) -
(\ref{cut3:bbgammagamma}) are shown in 
Table~\ref{tab:bbgammagamma:cutbycut}. Note that the cut on the $p_T$ of the hard jet is rather conservative since a larger cut such as $p_T(j_h) > 300$ GeV would reduce the background significantly while preserving most of the signal.
The cross section for the 
irreducible background after the cuts of Eq.~(\ref{cut1:bbgammagamma}) and Eq.~(\ref{cut2:bbgammagamma}) agree with those in
Refs.~\cite{Baur:2003gp,Contino:2012xk} to
within $\mathcal{O}(15\%)$. 
Given this 
agreement we use the efficiencies for the
reducible (subleading) 
backgrounds from Refs.~\cite{Baur:2003gp,Contino:2012xk} and assume
the same efficiency for the cut on the $p_T$ of the hard jet as in the
irreducible background.  We use the resulting efficiencies to estimate the 2 and
5$\sigma$ reach for this channel at the LHC
with $\sqrt{s}=14$ TeV for 300 fb$^{-1}$ of integrated luminosity  
and the results are shown in Fig.~\ref{fig:bbgammagamma}(a). For comparison the current bound in the degenerate bidoublet model as obtained from single
EW production is 
also shown as the blue dotted curve. In Fig.~\ref{fig:bbgammagamma}(b) we show
the confidence level as a function of the heavy quark mass for
different values of total integrated luminosity when $s_R$ is
fixed to the current upper bound.  We see that with an integrated
luminosity of $300$ fb$^{-1}$ we can obtain a 2 (5)$\sigma$
measurement up to 1.4 (1.0) TeV.

\begin{table}[t]
\begin{tabular}{|l|c|c|c|}\hline
cut & $\epsilon_{800}$ & $\epsilon_{1600}$ & $ \epsilon_{\mathrm{irred.}}$  \\\hline
Eq.(\ref{cut1:bbgammagamma}) & 0.14 & 0.087 & 0.00023 \\
Eq.(\ref{cut2:bbgammagamma}) & 0.76 & 0.7 & 0.13 \\
Eq.(\ref{cut3:bbgammagamma}) & 0.99 & 1 & 0.011 \\
\hline
\end{tabular}\hspace{0.6cm}
\caption{Efficiencies for the signal (with
$M=800$ and $1600$ GeV) 
and irreducible background in the $b\bar{b} \gamma \gamma$
channel for the various optimization cuts listed in
Eqs.~(\ref{cut1:bbgammagamma}) - (\ref{cut3:bbgammagamma}). 
}\label{tab:bbgammagamma:cutbycut}
\end{table}

\subsection{$b\bar{b}b\bar{b}$ channel}

Next we turn our attention to the channel with the largest
branching fraction, the one in which both Higgs bosons decay into $b\bar{b}$. 
\textit{A priori} a channel with only jets in the
final state is extremely challenging due to the immense QCD
background. A realistic determination of 
the feasibility of this multijet channel would require computing resources and
data-driven methods that are beyond the scope of this analysis. 
Hence we can only get a rough estimate of the LHC reach for this
channel. In this study we aim to point out the unique features of this
final state that will enhance signal sensitivity over the large
backgrounds. These unique features include a very hard jet, four $b$
quarks that reconstruct two Higgs bosons and the reconstructed heavy
quark. We believe that our results
are sufficiently promising to warrant a more detailed experimental study.

We have generated the following two background processes: the
irreducible $b\bar{b}b\bar{b}+1j$ and the $b\bar{b}+3j$
backgrounds. The corresponding cross 
sections are shown in Table~\ref{tab:bkgxsec}. The huge background
reduction resulting from the requirement of four b-tagged jets (we are
assuming a fake tag rate of 1 \%) allows us to neglect pure multijet
QCD backgrounds. 

\begin{figure}[htb]
{
\includegraphics[width=0.45\textwidth, clip=true]{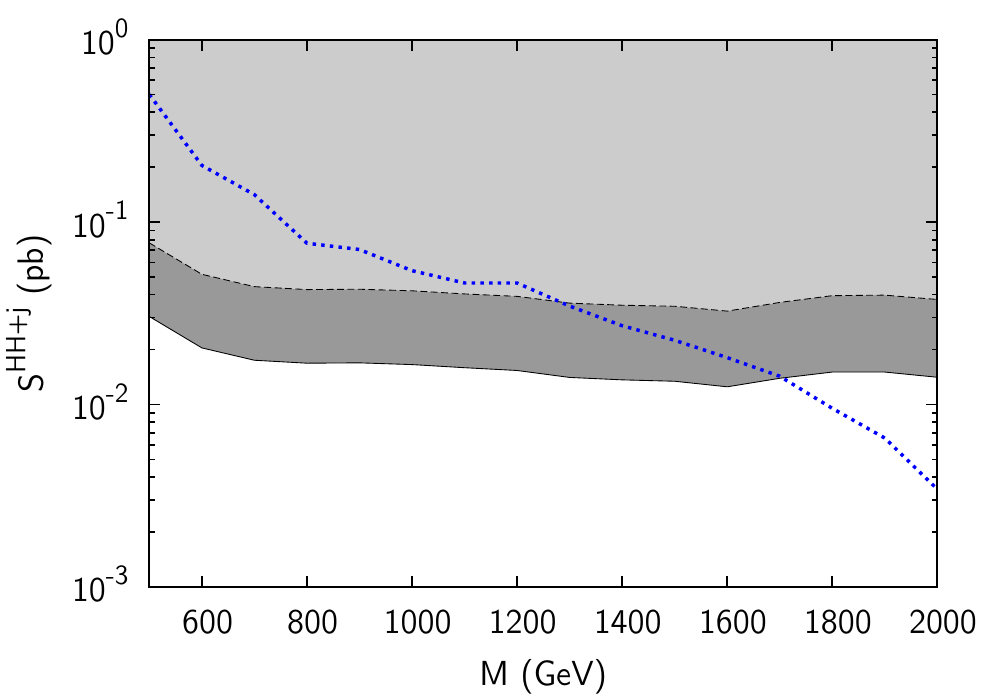}
\hfil
\includegraphics[width=0.45\textwidth, clip=true]{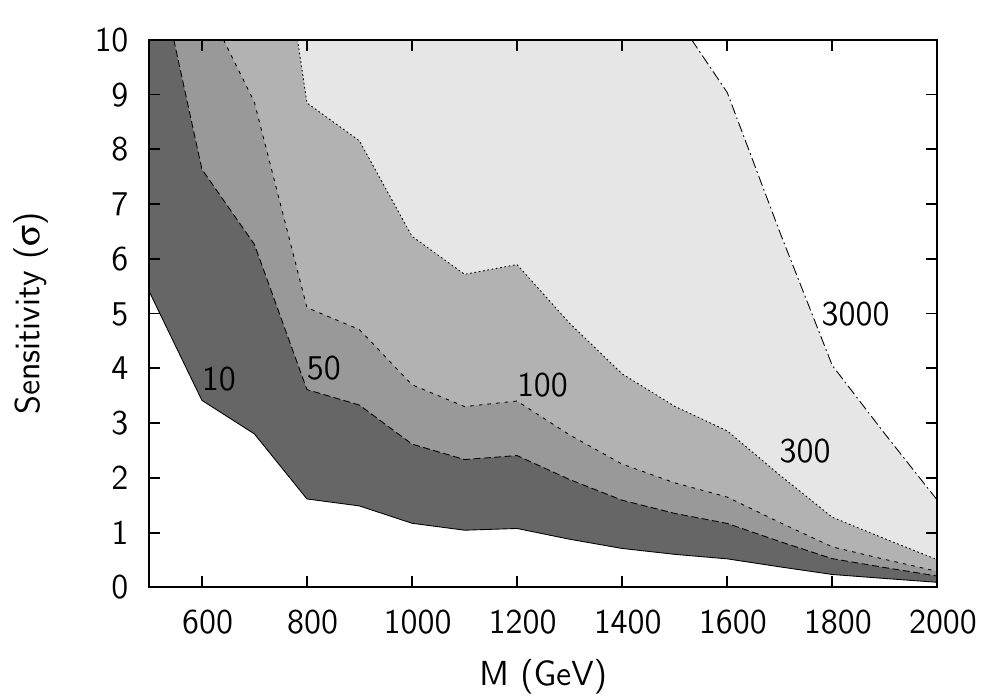}
}
\caption{
Same as Fig.~\ref{fig:sensitivityWF} but for the $HHj$ channel with
  $b\bar b b\bar b j$ final state.}  
\label{fig:bbbb}
\end{figure}

We propose the following cuts:
\begin{itemize}
\item At least 5 jets, four of which must be tagged as b-jets.
\item $p_T(j_h)> 300$ GeV.
\item $|m_{jj}-m_H|<50$ GeV (for both pairs of b-jets).
\end{itemize}
The very large $b\bar{b}+3j$ cross section (see
Table~\ref{tab:bkgxsec}) is greatly reduced due to the requirement of
four b-tags, given the $1\%$ mistag rate we consider. This huge
reduction makes it very challenging to generate enough statistics to
reasonably estimate the efficiency of the remaining cuts. In order to
estimate this efficiency we have \textit{not} imposed the requirement of
the four b-tags but rescaled the corresponding cross sections with the
factors resulting from the b-tagging efficiency ($\epsilon_{4b}(bbbbj)\approx 0.25$) or mistagging rate ($\epsilon_{4b}(bbjjj)\approx
1.5\times 10^{-4}$). The remaining cuts on the transverse momentum of the hard jet and the Higgs mass reconstruction are then implemented.  
The pairs of jets used to reconstruct the Higgs boson are selected from the four
sub-leading jets, as in the signal the hardest jet is typically the
one from the decay of $U_H$ and is not a $b$ quark.
The two jets that reconstruct the Higgs mass closest to $m_H = 125$
GeV form the first pair while the 
remaining two jets reconstruct the other Higgs boson.
Once the two Higgs candidates are selected we construct 
the invariant mass of one Higgs boson and the leading jet and 
require that at least one 
of these invariant masses be in the region
\begin{equation}
|m_{Hj_h}-M|<400\mbox{ GeV}.
\end{equation}
We show the corresponding efficiencies after all cuts for the signal
and the two backgrounds we have considered in Table~\ref{tab:bbbb} and
a summary of the results in Fig.~\ref{fig:bbbb}. Keeping in mind the
inherent lack of precision in the estimation of the backgrounds for 
this process we see that this channel is potentially even more promising than
the $b\bar{b}\gamma\gamma$ one. A 2 (5)$\sigma$ sensitivity could be
obtained 
for masses up to 1.7 (1.3) TeV
with 300 fb$^{-1}$ integrated luminosity. 

\begin{table}[ht]
\begin{tabular}{|l|c|c|c|c|c|}\hline
M (GeV) & $\sigma_s$ (fb) & $\epsilon_{s}$ & $\epsilon_{4b+j}$ & $ \epsilon_{2b+3j}$   & bk
events/100 fb$^{-1}$ \\\hline
500 & 125 & 0.033 & 0.0051 & 0.0029 & 456\\
1000 & 12.6 & 0.057 & 0.005 & 0.003 & 412\\
1500  & 5.6 & 0.03 & 0.0008 & 0.0005 & 71\\
\hline
\end{tabular}\hspace{0.6cm}
\caption{Cross sections (in fb) for the signal ($b\bar{b}b\bar{b}j$ channel) and efficiencies for the signal and main
backgrounds after all the cuts listed in the text for different values of heavy quark mass, $M$. The total 
  number of background events with 100 fb$^{-1}$ of integrated
  luminosity is also shown in the last column. The efficiency due to
  b-tagging as described in the text is not included in this table but
  it is used to compute the number of background events. 
}\label{tab:bbbb}
\end{table}

\section{Discussion\label{sect:conclusions}}

New vector-like quarks can mix sizeably with first generation SM
quarks without conflicting with current experimental constraints. 
The underlying mechanism that allows these new vector-like quarks to
evade the very stringent indirect constraints requires more than one
new quark so that delicate cancellations can take place. These
cancellations can be naturally enforced by symmetries that typically
imply the new quarks to be degenerate and to have unique decays (with
100$\%$ branching fractions) into a SM quark and either an EW gauge
boson or the Higgs boson. Channels involving new heavy quarks with decays into
EW gauge bosons are currently being searched for at the LHC and have
become the most stringent probes of these new quarks. The equivalence
theorem however guarantees that if there are channels involving decays
into EW gauge bosons then the channels with decays into the
Higgs boson must also be present. Direct searches of the channels involving decays into the
Higgs boson are therefore a crucial ingredient to fully disentangle
the mechanism underlying the protection of the SM quark couplings 
in these new physics scenarios.

We have used as a benchmark a well motivated model that allows a
large mixing of new vector-like quarks with first generation SM
quarks, the so called degenerate bidoublet model. This model involves
four new (essentially degenerate) quarks, each one decaying, with
100$\%$ branching fraction, into an up quark and a $Z$, $H$, $W^+$ and
$W^-$, respectively. After discussing the current direct and indirect
constraints on the model we have described the main production
processes that involve the new quark with decays into a Higgs boson.
We have proposed new searches that are sensitive to this type
of new quark and have discussed the potential reach at the LHC.
 
The two most promising Higgs channels are vector boson Higgs fusion
and associated production. The former channel results 
in a final state with a Higgs, an EW gauge boson and a
hard jet. The charged current channel is particularly promising, with a
reach of up to 2 TeV for the heavy quark mass with 300 fb$^{-1}$ of integrated
luminosity. In the associated production channel there are two Higgs
bosons and a hard jet in the final state but it suffers from a smaller
cross section. In the case where one Higgs boson decays into $b$ quarks and the other into photons we have shown that
a 5$\sigma$
sensitivity can be reached with 300 fb$^{-1}$ of integrated
luminosity for a mass of the heavy quark below 1 TeV. 
We have also considered the case in
which both Higgs bosons decay into $b\bar{b}$. Despite the
all hadronic final state, the large number (four) of $b$ quarks and a
very hard extra jet are powerful enough discriminators to make this a
very promising channel as well, with a reach of up to 1.3 TeV for the heavy quark mass with 300 fb$^{-1}$. 

A measurement of new vector-like quarks that decay into the
Higgs boson and a SM light quark would represent a crucial direct test
of the underlying mechanism that protects the SM quark gauge couplings
from large corrections. Our results show that this measurement is
feasible at the LHC for a large portion of the currently allowed
parameter space, in several different channels simultaneously.

\begin{acknowledgments}
This work has been partially supported by MICINN projects
FPA2006-05294 and FPA2010-17915, through the FPU program and by Junta
de Andaluc\'{\i}a projects FQM 101, FQM 03048 and FQM 6552. AA is supported by the United States National Science Foundation under grant PHY-0854889.
\end{acknowledgments}

\begin{center}
\end{center}
\vspace{-1.5cm}
\bibliographystyle{apsrev}
\bibliography{myrefs}


\end{document}